\documentclass{aa}       
\usepackage{graphicx}
\usepackage{amssymb}
\usepackage{amsmath}
\usepackage{longtable}
\usepackage{supertabular}
\usepackage{natbib}
\bibpunct{(}{)}{;}{a}{}{,}

\begin{document}

\title{3D simulations of wind-jet interaction in massive X-ray binaries}

\author{M. Perucho\inst{1} \and
        V. Bosch-Ramon \inst{2} \and
        D. Khangulyan \inst{3,2}
    }

\authorrunning{Perucho, Bosch-Ramon \& Khangulyan}

\titlerunning{3D simulations of wind-jet interaction in massive X-ray binaries}

\institute{Dept. d'Astronomia i Astrof\'{\i}sica, Universitat de Val\`encia, C/ Dr. Moliner 50, 46100, Burjassot (Val\`encia), Spain; Manel.Perucho@uv.es 
\and
Max Planck Institut f\"ur Kernphysik, Saupfercheckweg 1, Heidelberg 69117, Germany; vbosch@mpi-hd.mpg.de
\and
Institute of Space and Astronautical Science/JAXA, 3-1-1 Yoshinodai, Sagamihara, Kanagawa 229-8510, Japan; Dmitry.Khangulyan@mpi-hd.mpg.de
}

\offprints{M. Perucho, \email{manel.perucho@uv.es}}

\date{Received <date> / Accepted <date>}

\abstract
{High-mass microquasars may produce jets that will strongly interact with surrounding stellar winds on binary system spatial scales.}  
{We study the dynamics of the collision between a mildly relativistic hydrodynamical
jet of supersonic nature and the wind of an OB star.} 
{We performed numerical 3D simulations of jets that
cross the stellar wind with the code \textit{Ratpenat}.}
{The jet head generates a strong shock in the wind, and
strong recollimation shocks occur due to the initial overpressure of the jet
with its environment. These shocks can accelerate particles up to TeV energies and produce gamma-rays. 
The recollimation shock also strengthens jet asymmetric Kelvin-Helmholtz instabilities
produced in the wind/jet contact discontinuity. This can lead to jet disruption even for jet powers of several times
$10^{36}$~erg~s$^{-1}$.}{High-mass microquasar jets likely suffer 
a strong recollimation shock that can be a site of particle acceleration up to very high energies, 
but also eventually lead to the disruption of the jet.} 
\keywords{X-rays: binaries--ISM: jets and outflows--Stars: winds, outflows--Radiation mechanisms: non-thermal}

\maketitle

\section{Introduction} \label{intro}

Jets of X-ray binaries (microquasars) are produced close to the compact object (black hole or neutron star) via ejection of
material accreted from the stellar companion.  Jet synchrotron emission was extensively observed (Rib\'o 2005), and its
phenomenological properties and connections with other energy bands were thouroughly analyzed (Fender et al. 2004). The occurrence of
collisionless shocks in microquasar jets can lead to efficient particle acceleration (Rieger et al. 2007) and non-thermal
emission of synchrotron and inverse Compton origin and, possibly, from proton-proton collisions (see Bosch-Ramon \&
Khangulyan 2009 and references therein). To understand the dynamics of these jets, numerical simulations of their propagation
were performed (Peter \& Eichler 1995; Vel\'azquez \& Raga 2000; Perucho \& Bosch-Ramon 2008;  Bordas et al. 2009)
together with several analytical treatments (Heinz \& Sunyaev 2002; Heinz 2002; Kaiser et al. 2004; Nalewajko \& Sikora 2009;
Araudo et al. 2009). Most of these works were valid for microquasars in general, although jets of high-mass systems require a
specific treatment due to the strong stellar wind.

Perucho \& Bosch-Ramon (2008) --PBR08 from now on-- studied how the strong wind of an OB star can influence the jet dynamics
at scales similar to the orbital separation ($\sim 0.2$~AU). Simulations in two dimensions of a hydrodynamical jet
interacting with an homogeneous (i.e. not clumpy) stellar wind were performed in cylindrical and slab symmetries.   The
results showed that a strong recollimation shock is likely to occur at jet heights $\sim 10^{12}$~cm, which could lead to
efficient particle acceleration and gamma-ray emission, which was detected in several high-mass X-ray binaries (HMXB) (LS~5039,
LS~I~+61~303, Cygnus~X-1, Cygnus~X-3; Aharonian et al. 2005, Albert et al. 2006, Albert et al. 2007, Tavani et al. 2009). It
was also found that jet disruption could occur for jet kinetic luminosities as high as $L_{\rm j}\sim 10^{36}$~erg~s$^{-1}$
because of jet instabilities produced by the strong and asymmetric wind impact. Such a $L_{\rm j}$-value is $\sim 0.1-1$\% of
the Eddington luminosity, which is typical for an X-ray binary with persistent jets (see Fig.~1 in Fender et al. 2003). All this means
that the role of the stellar wind in high-mass microquasars (HMMQ) is not only to feed accretion, but it may be also relevant
for particle acceleration within the binary system (i.e. in strong recollimation shocks) and a detectable
jet at larger spatial scales. However, although the assumptions in PBR08 of a hydrodynamical jet and a homogeneous wind seem
reasonable at the involved jet heights of $\sim 10^5$~$R_{\rm Sch}$ (Sikora et al. 2005) and under moderate clumping (see
Owocki \& Cohen 2006), the assumption of 2-dimensional symmetry is far less realistic. Therefore, 3-dimensional (3D)
simulations of the wind impact on the jet are necessary to understand the non-thermal phenomena occurring in HMMQ.

In this Letter, we present the results of 3D simulations of hydrodynamical jets interacting with a typical OB star wind. We
find that even jets with $L_{\rm j}$ as high as several times $10^{36}$~erg~s$^{-1}$ may be disrupted, since the lower
wind-jet momentum transfer due to wind sidewards escape is balanced by enhanced development of disruptive instabilities.
The simulations also show a strong recollimation shock that could efficiently accelerate particles. A deeper treatment of the
radiative counterpart of the simulations presented here is in preparation (Bosch-Ramon, Khangulyan \& Perucho, in
preparation).

\section{Simulations}

We performed 3D simulations of two supersonic hydrodynamical jets with $L_{\rm j}=10^{35}$ (Jet 1) and
$10^{37}$~erg~s${}^{-1}$ (Jet 2). The medium is an isotropic wind (as seen from the star) of a mass-loss rate of 
$10^{-6}$~$M_\odot$~yr${}^{-1}$ and a constant velocity of $2\times 10^8$~cm~s${}^{-1}$ (PBR08).  The companion star is assumed to
be located at $R_{\rm orb}=2\times 10^{12}$~cm from the base of the jet in a direction perpendicular to the jet axis. Both
jets are injected at a distance to the compact object of $z_0=6\times 10^{10}$~cm with an initial jet radius 
$R_{\rm j}=6\times
10^9$~cm. Their initial velocity, temperature and Mach number are $0.55\,c$ ($1.7\times10^{10}$~cm~s${}^{-1}$), 
$10^{10}$~{K} and $17$, respectively. The initial densities and pressures for the jets 1 and 2 are $\rho_1=0.088\,\rho_{\rm
w}$ and $\rho_2=8.8\,\rho_{\rm w}$, and $P_{\rm j,1}=71$~erg~cm${}^{-3}$ and $P_{\rm j,2}=7.1\times10^3$~erg~cm${}^{-3}$,
respectively; $\rho_{\rm w}=3\times 10^{-15}$~g~cm${}^{-3}$ is the wind density at $z_0$. We used the coordinate $z$ for the
direction of the initial propagation of the jet, $x$ for the direction connecting the jet base and the star, and $y$ for the
direction perpendicular to $z$ and $x$. We assumed that the magnetic field has no dynamical influence in the evolution of the
jet, the wind is continuous and homogeneous and the compact object is at the same orbital position during the simulation
time, $\sim 10^3$~s, much smaller than the orbital period.

We used a finite-difference code named \textit{Ratpenat}, which solves the equations of relativistic hydrodynamics in
three dimensions written in conservation form using high-resolution-shock-capturing methods. \textit{Ratpenat} was
parallelized with a hybrid scheme with both parallel processes (MPI) and parallel threads (OpenMP) inside each process (see
Perucho et al. 2009). The simulations were performed in Mare Nostrum, at the Barcelona Supercomputing Centre (BSC) with
up to 128 processors. The numerical grid box expands transversely 20~$R_{\rm j}$ on each side of the axis, which amounts to a total of
40~$R_{\rm j}$, and 320~$R_{\rm j}$ 
along the axis. The numerical resolution of the simulation is of four
cells per initial jet radius. This means that the final box is $160\times160\times 1280$ cells. An extended grid is used in
the transversal direction, composed by 80 cells with an increasing size, which brings the outer boundary 80~$R_{\rm j}$ farther
from the axis. The resolution is low at $z_0$ due to the amount of computational time needed to perform the simulations, but
because jets initially expand, the effective resolution at the $z$ of interest ($\sim 10^{12}$~cm) in the main grid is $\sim
16$ cells per $R_{\rm j}$. To check possible resolution problems, a double resolution simulation of Jet 2 was performed
up to one half of the jet final distance (which required 10 times longer). The accumulated error in this first portion of the
simulation is negligible for our purposes in all the hydrodynamical variables.

\section{Results}

\begin{figure}[!h]
     \centering
  \includegraphics[clip,angle=0,width=\columnwidth]{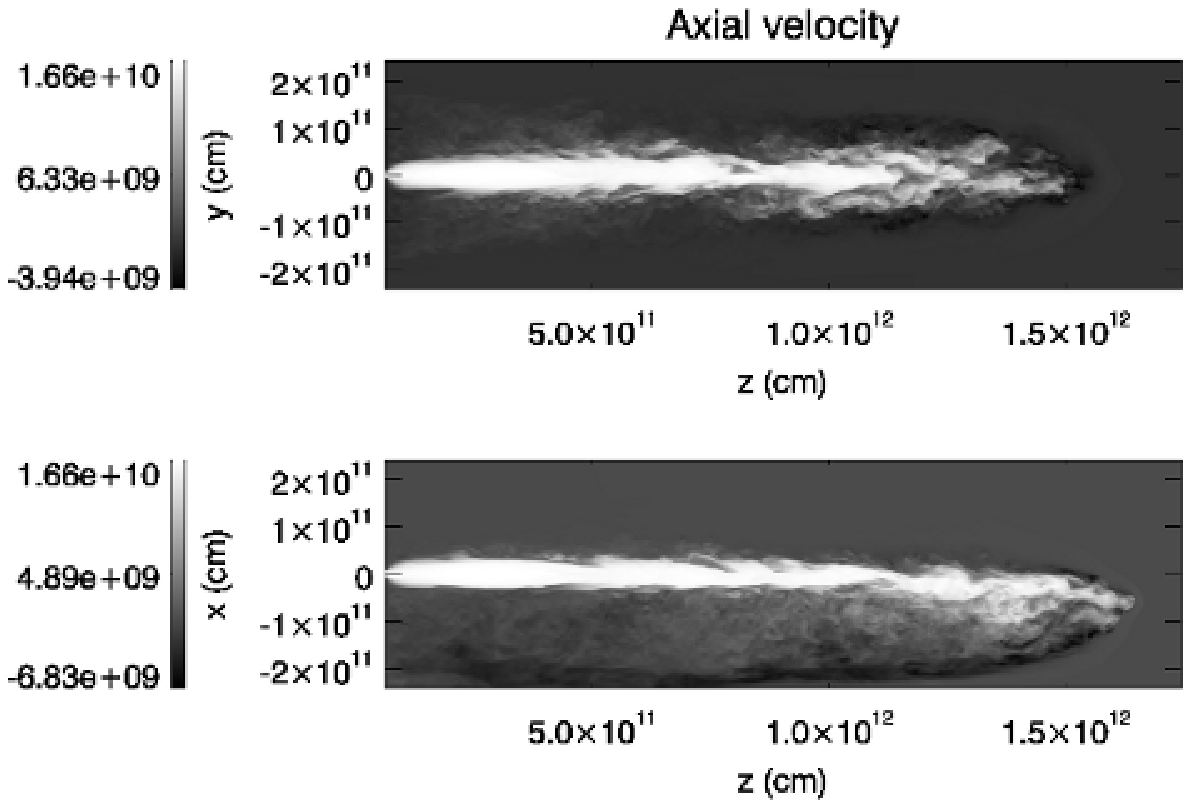}
  \includegraphics[clip,angle=0,width=\columnwidth]{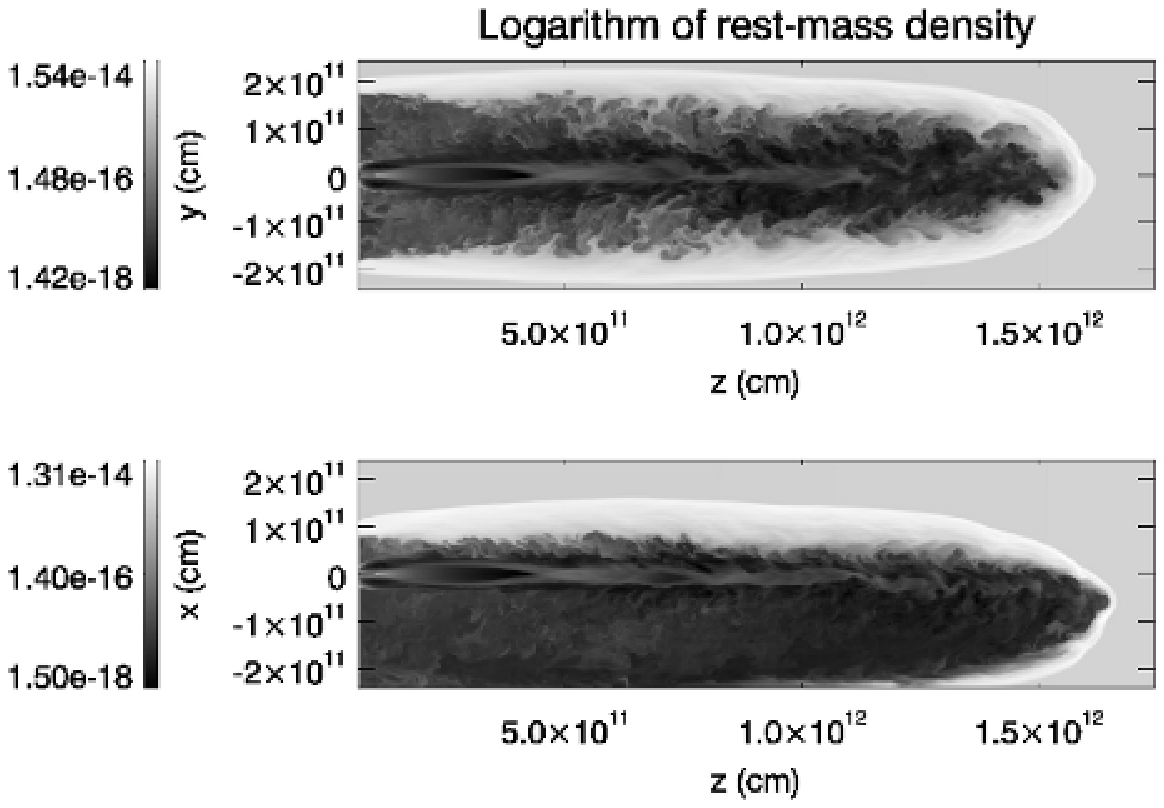}
  \caption{Jet 1, cuts of axial velocity (in $\rm{cm\,s^{-1}}$) and rest-mass 
  density (in $\rm{g\,cm^{-3}}$) along the propagation axis. The upper two panels show a cut perpendicular to the 
  star-jet plane, whereas the bottom panels show a cut in this plane. The scale of the figure respects 
  the proportion with Fig.~\ref{fig:maps2}.}
  \label{fig:maps1}
  \end{figure}

Jet 1 propagates up to $z\leq 1.6\times10^{12}$~cm after $\simeq 1.25\times10^3$~s. Just after injection, the difference
between the velocity of the jet head and that of the backflow generates a low density region around the jet base, and the
backflow material fills it very slowly. The jet expands initially in this region, mixing with some of the cocoon
material and finally generating a thick shear layer with positive velocities. The backflow interacts strongly with this outer
region, generating instabilities that grow in the shear layer. These instabilities are quite asymmetric due to the nature of
the cocoon in the plane of impact of the wind (see Fig.~\ref{fig:maps1}). After expansion, the central region of the jet
becomes underpressured with respect to its surroundings and recollimates until the formation of a reconfinement shock. Thism shock is quasi-steady, i.e. propagates very slowly from $z_{\rm s}\simeq 2\times10^{11}$~cm to $4\times10^{11}$~cm, as
the pressure in the cocoon drops (see next section). Downstream of the reconfinement shock, the instabilities, which were
growing in the shear layer, propagate to the whole section of the jet as the internal jet flow is decelerated and becomes
more sensitive to perturbations. This process ends up in the mixing and deceleration of the jet flow at $z\geq 10^{12}$~cm. 

  \begin{figure*}[!t]
     \centering
  \includegraphics[clip,angle=0,width=\textwidth]{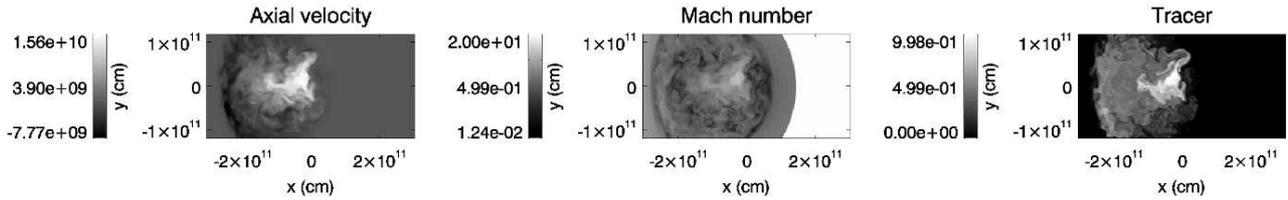}
  \caption{Transversal cuts for the axial velocity, Mach number and tracer in Jet 1 at $z\simeq 1.3\times10^{12}\,\rm{cm}$. The Mach number is saturated for values higher than 20 (the wind).}
  \label{fig:mapst1}
  \end{figure*} 

At the end of the simulation, the velocity of the bow shock at jet head is $\simeq 3\times10^{8}$~cm~s${}^{-1}$. The fit for
the bow-shock advance velocity results in a fast decrease with time: $v_{\rm bs}\propto t^{-0.6}$. Figure~\ref{fig:maps1}
shows two axial cuts of Jet 1 with axial velocity (two upper panels) and rest-mass density (two lower panels), along the plane perpendicular to the star-jet plane (upper) and in this plane (lower) at the last snapshot. The deviation caused by the wind thrust can be
observed in the plane $xz$, which is the plane formed by the optical star and the jet. The low velocity of the jet head at
the end of the simulation along with its final (destabilized) structure implies that the jet will be disrupted and will not
propagate out of the binary system as a supersonic and collimated flow. Figure~\ref{fig:mapst1} shows transversal cuts of the
axial velocity, Mach number and tracer\footnote{The tracer, $f=\left[0,1\right]$, is a variable that indicates the
composition of the fluid in a given cell (see e.g., Perucho et al. 2005), with a value of 0 corresponding to pure wind
material, 1 to pure jet material, and any value between 0 and 1 indicating the relative amount of jet material in a cell where
mixing has occurred.} at $z\simeq 1.3\times10^{12}$~cm. In this figure we can see the deformation of the bow shock caused by
the wind thrust. The jet has been entrained up to its axis by the wind material ($f_{max}<1$). The maximum velocity in the
jet fluid is still relatively fast ($\simeq 1.5\times10^{10}$~cm~s${}^{-1}$) despite the irregular morphology and mixing, but
the average Mach number is close to one.  

\begin{figure}[!h]
     \centering
  \includegraphics[clip,angle=0,width=\columnwidth]{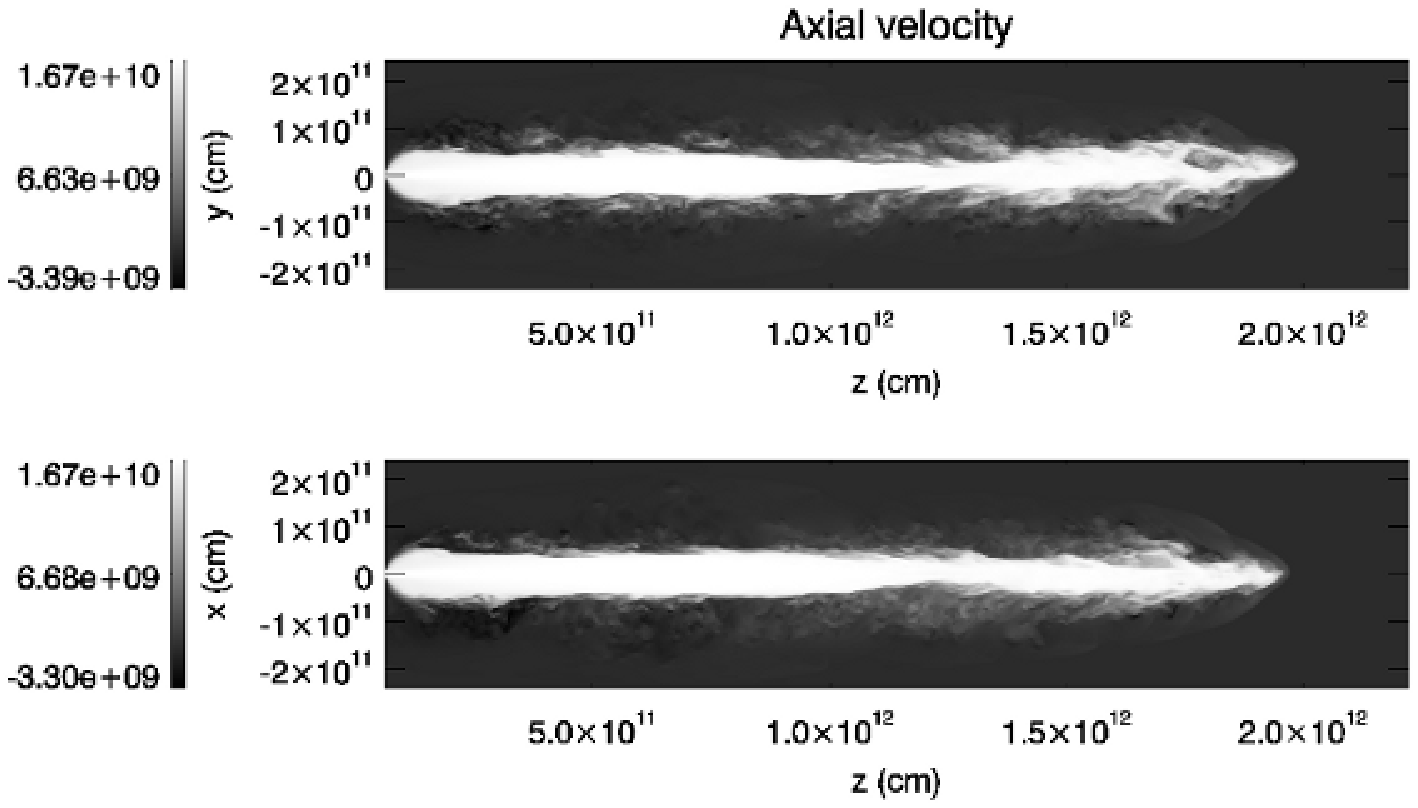}
  \includegraphics[clip,angle=0,width=\columnwidth]{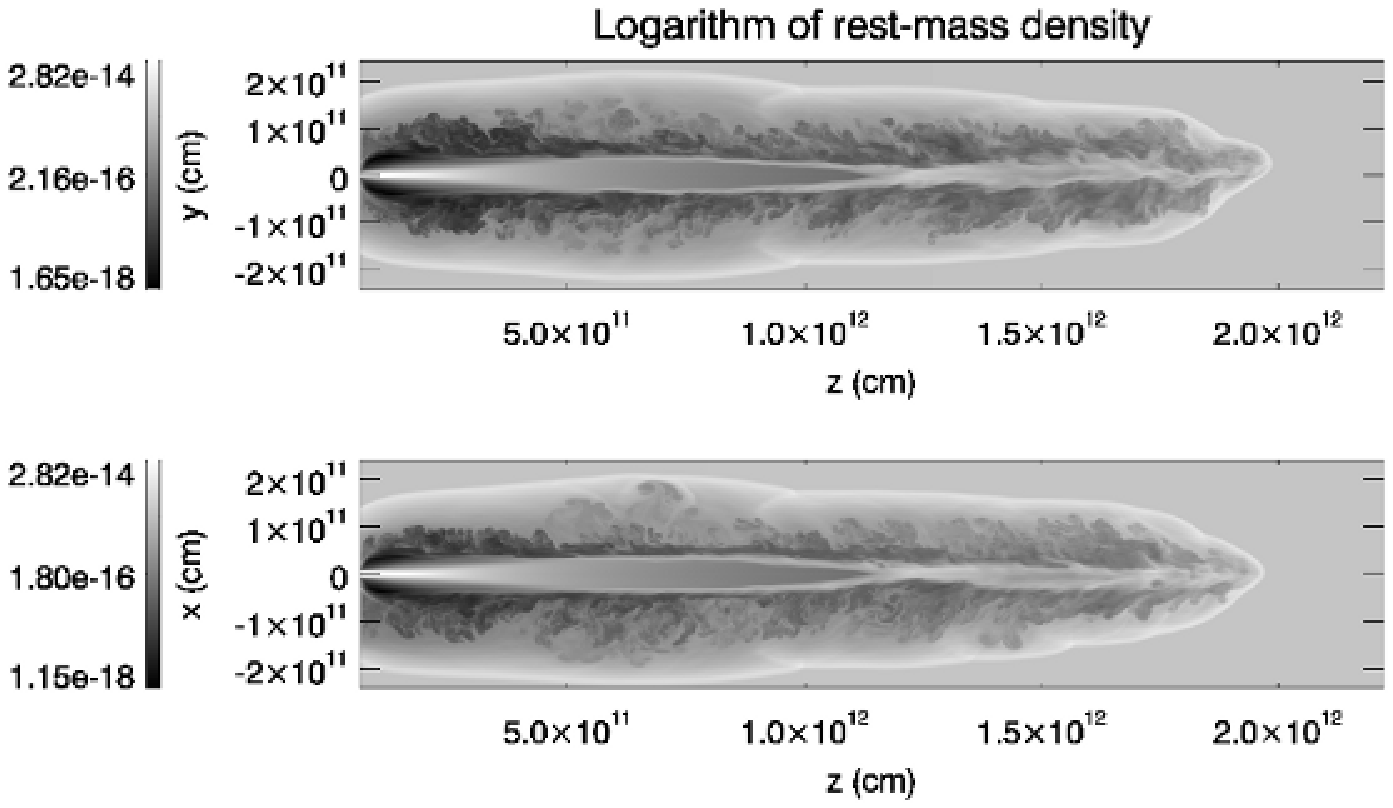}
  \caption{Same plots as in Fig.~\ref{fig:maps1}, for Jet 2.}
  \label{fig:maps2}
  \end{figure}

\begin{figure*}[!t]
     \centering
  \includegraphics[clip,angle=0,width=\textwidth]{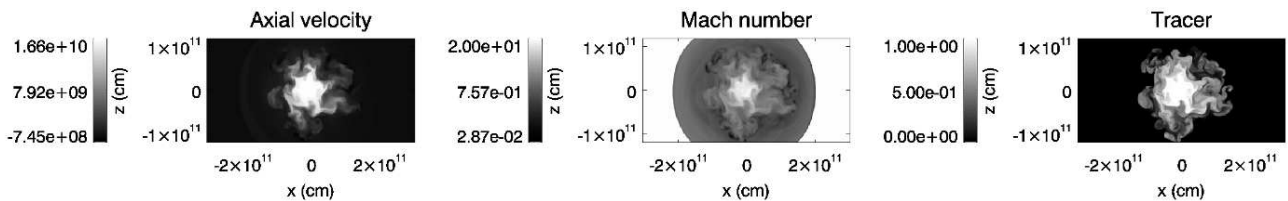}
  \caption{Same as in Fig.~\ref{fig:mapst1} for Jet 2 at $z\simeq 1.5\times10^{12}\,\rm{cm}$.The Mach number is saturated for values higher than 20 (the wind).}
  \label{fig:mapst2}
  \end{figure*}

The evolution of Jet 2 is similar to that of Jet 1 from a qualitative point of view. It propagates up to $z\sim
2\times10^{12}$~cm in $\simeq 210$~s. The jet expands more at the base because it is initially denser (more overpressured),
the velocity of the jet head is faster than in Jet 1, and the cocoon at this $z$ is consequently filled more slowly.
Therefore, the reconfinement shock is stronger and occurs at larger $z$  (see next section). The location of the shock
changes with time from $z\simeq 6\times 10^{11}$~cm to $10^{12}$~cm (see Fig.~\ref{fig:maps2}). The effect of the wind in the
direction of the jet propagation is small, as seen in Figs.~\ref{fig:maps2} and \ref{fig:mapst2}. In the latter, the
structure of the bow shock at $z\simeq 1.3\times10^{12}$~cm is observed to be more symmetric than that in Jet 1
(Fig.~\ref{fig:mapst1}). Figure~\ref{fig:mapst2} shows that the jet core is unmixed ($f=1$) and that the flow velocity is still
as high as that in the injection point. At the end of the simulation, the velocity of the bow shock is $\simeq
8.4\times10^{9}$~cm~s${}^{-1}$, which is close to its initial speed ($\simeq 9.\times10^{9}$~cm~s${}^{-1}$). However, the
differences in the the physical conditions in the cocoon on both sides of the jet cause the development of asymmetric
Kelvin-Helmholtz instabilities in the shear layer. Their propagation to the whole jet after the reconfinement shock triggers
helical motions and distortions in the flow.

At $z\ga R_{\rm orb}$, we expect the density of the stellar wind to be $\propto z^{-2}$. This implies a pressure drop and a
jet adiabatic expansion. The helical instability triggered at $z\la R_{\rm orb}$ will then develop in these changing
conditions and may slow its growth (Hardee 2006). How much it will affect the stability of the jet flow in an
expanding jet yet remains to be studied. However, since the instability reaches nonlinear amplitudes, the process is irreversible
and we do not expect it to disappear. Therefore $L_{\rm j}=10^{37}$~erg~s${}^{-1}$ seems to be close to the minimum jet
power to propagate out of the binary region without disruption.

\section{Discussion}

The performed simulations show that a recollimation shock likely forms when the jet is crossing the wind. As mentioned, this
shock can trigger particle acceleration, and could also enhance instabilities that would destroy the jet. However, the
jet-wind interaction simulated here is time-dependent, and the study of the stationary case calls for a specific simulation.
Still, it is worthwhile to estimate here analytically whether the shock remains inside the binary region when the jet head is
already outside. This will depend on whether the wind ram pressure can substitute the cocoon pressure to keep the
recollimation shock inside the system. 

From Falle (1991), we know that $z_{\rm s}\propto P_{\rm{c}}^{-1/2}$, with $P_{\rm{c}}$ the pressure of the cocoon, if the
jet power and mass flux are kept constant. Thus we need to know how this pressure changes with time. Following Scheck et al.
(2002) and Perucho \& Mart\'{\i} (2007), we find that $P_{\rm{c}}\propto t^{-1-\alpha/2}$ for a homogeneous ambient medium,
where $v_{\rm{bs}}\propto t^{\alpha}$; and $P_{\rm{c}}\propto t^{-2}$ when the ambient medium density decreases as
$\rho_{\rm{a}}\propto z^{-2}$. This implies $z_{\rm s}\propto t^{1/2+\alpha/4}$ and $\propto t$, respectively. This result
can be tested using our simulation, in which the ambient medium is still roughly homogeneous (at $z\la R_{\rm orb}$). For Jet 2 the recollimation shock forms at $z=6\times 10^{11}$~cm at $t_1\simeq 61$~s. At $t_2\simeq 210$~s, taking into account
that the velocity is basically constant ($\alpha=0$), $z_{\rm s}(t_2)/z_{\rm s}(t_1)=(t_2/t_1)^{1/2}$, which results in
$z_{\rm s}(t_2)=1.1\times 10^{12}$~cm, in agreement with the simulation. We can then extrapolate to find out
the time ($t_3$) at which the shock would reach $z=2\times 10^{12}$~cm. Taking $t_2=210$~s, and $z_{\rm s}(t_2)=1.1 \times
10^{12}$~cm we obtain for $\rho_{\rm{a}}\propto z^{-2}$: $t_3 \simeq 2\,t_2 \simeq 400$~s. 

We now calculate the time at which the pressure of the cocoon will fall below the pressure of the shocked wind. This is
the time from which the recollimation shock will be determined by the latter, which is $P_{\rm{w}}\sim 10^2$~erg~cm${}^{-3}$
(see PBR08). Taking into account that the mean pressure in the cocoon is $P_{\rm{c}}\simeq 10^3\,\rm{erg\,cm^{-3}}$ at
$t_2=210\,\rm{s}$, pressure equilibrium with the shocked wind will be reached at $t\simeq
\sqrt{P_{\rm{c,0}}/P_{\rm{w}}}\,\,t_2\simeq 3\,t_2$. This means that Jet 2 is around the limit to keep the recollimation shock
inside the system by the wind ram pressure alone, under the simulated conditions. For hotter or denser jets, the
recollimation shock can move out of the region of interest at a finite time, whereas for colder or lighter jets, this shock
will stay inside the binary system. This allows for continuous production of energetic emission, but puts the jet in danger of
disruption. An X-ray binary in which a disruption of the jet could be taking place is LS 5039 (Mold\'on et al. 2008),
but higher angular resolution observations are required for a proper probe of the jet-wind interaction region.

The scenario considered in this work could take place in several HMXB in the Galaxy. The luminosity function derived by Grimm
et al. (2003) predicts $\sim 3$ HMXBs with $L_X=10^{35}\rm{erg/s}$. Following Fender et al. (2005), an HMXB with a
$10\,M_\odot$ black hole could produce a jet with a kinetic power between $10^{35}$ and $10^{38}\rm{erg/s}$, which is in the
range of the simulations performed here. Although Grimm et al. (2003) do not offer a specific prediction for $L_X \leq
10^{35}\rm{erg/s}$, extrapolating the given luminosity function we deduce that there is room for $\sim 10$ sources in
our Galaxy whose jets could be disrupted by the stellar wind. These objects may be bright at high energies, but faint or even
quiet in radio.

Several improvements are required to make 3D HMMQ jet simulations more realistic: to study a stationary case at the binary
scales; to perform detailed calculations of the radiation produced in the jet-wind shocks; to introduce an inhomogeneous or
clumpy stellar wind; to follow the jet head up to $z\gg R_{\rm orb}$ to study the effects of the decreasing wind density; and
to account for the effect of the magnetic field on the dynamics of these jets, since the configuration of the magnetic field
in the jet affects its stability (Hardee 2007, Mizuno et al. 2007). All this is ongoing work to be presented elsewhere.

\begin{acknowledgements}
MP acknowledges support from a ``Juan de la Cierva'' contract of the Spanish ``Ministerio de Ciencia y Tecnolog\'{\i}a'' and
from the Spanish ``Ministerio de Educaci\'on y Ciencia'' through grants AYA2007-67627-C03-01, CSD2007-00050 and
AYA2007-67752-C03-02. The authors acknowledge the \emph{Barcelona Supercomputing Center} for support and the ``Red Espa\~nola
de Supercomputaci\'on'' for the computing time allocated for this project. V.B-R. acknowledges support by the Ministerio de
Educaci\'on y Ciencia (Spain) under grant AYA 2007-68034-C03-01, FEDER funds. V.B-R. thanks Max Planck Institut fuer
Kernphysik for its kind hospitality and support.
\end{acknowledgements}

\end{document}